# Subspace projection method for unstructured searches with noisy quantum oracles using a signal-based quantum emulation device

Brian R. La Cour[1] · Corey I. Ostrove[1]

**Abstract** This paper describes a novel approach to solving unstructured search problems using a classical, signal-based emulation of a quantum computer. The classical nature of the representation allows one to perform subspace projections in addition to the usual unitary gate operations. Although bandwidth requirements will limit the scale of problems that can be solved by this method, it can nevertheless provide a significant computational advantage for problems of limited size. In particular, we find that, for the same number of noisy oracle calls, the proposed subspace projection method provides a higher probability of success for finding a solution than does an single application of Grover's algorithm on the same device.

## 1 Introduction

A great many problems can be described in terms of a general Boolean function $f : \{0, \ldots, 2^n - 1\} \mapsto \{0, 1\}$ mapping an $n$-bit integer to a single bit and for which $f(x) = 1$ indicates a solution [9]. The general problem is, then, to determine the number of solutions, if any, and their respective values. In the absence of any structural information regarding $f$, the general solution requires the evaluation of $f$ for all $2^n$

✉ Brian R. La Cour
blacour@arlut.utexas.edu

Corey I. Ostrove
costrove@arlut.utexas.edu

[1] Applied Research Laboratories, The University of Texas at Austin, P.O. Box 8029, Austin, TX 78713-8029, USA

possible inputs. Clearly, the complexity of this problem grows exponentially in the number of input bits.

Suppose that $f$ can be evaluated in a number of steps that scale only polynomially in $n$. The problem of determining whether a solution exists falls within the complexity class NP (Nondeterministic Polynomial), while finding the total number of solutions falls within the complexity class #P [2,10]. In this paper, we shall be concerned with the most general problem of determining the number and values of all solutions.

A quantum computer implementing Grover's search algorithm would be capable of a quadratic speedup in unstructured searches, but for problems that fall outside the complexity class P this approach is still inefficient [13,22]. In the absence of quantum error correction, decoherence and noise may further degrade performance [19,25–27]. Abrams and Lloyd have noted that a nonlinear quantum computer could solve such problems efficiently [3], but no such nonlinearity in quantum phenomena has been observed experimentally [6,20,29,30]. Along similar lines, Ohya and Volovich have suggested that a *chaotic* quantum computer would allow a solution in polynomial time [23,24], but no practical implementation of this scheme is yet known. Nuclear magnetic resonance (NMR) bulk-ensemble computing offers an alternative approach to solving unstructured search problems that is practically implementable for a few dozen qubits and capable of an exponential speedup [7,31] Quantum annealing has also been used to solve certain NP-complete problems that can be represented as quadratic binary optimization (QUBO) problems [11,12,15], but it is not known whether this approach is intrinsically more efficient than Grover's search algorithm or even standard classical methods [14]. Finally, Khatri and coworkers have developed a novel noise-based logic scheme to solve Boolean satisfiability problems using an orthogonal basis of independent classical noise processes [18].

Recently, a classical, signal-based scheme was proposed for emulating a gate-based quantum computer. Using a set of sinusoidal signals to represent individual qubits, a quadrature-modulated tonal (QMT) representation of the quantum state may be constructed that reproduces the Hilbert space structure of a multi-qubit quantum computer [17]. For a signal represented as a time-varying electronic voltage, standard analog electronic devices may be used to construct the quantum state, manipulate it through a sequence of gates, and perform a measurement resulting in a digital output. An analog emulation of this sort differs significantly from a traditional digital simulation in that it provides an explicit physical representation of states within the Hilbert space, and the operations upon them, thereby giving rise to a fundamentally different computational model. Since the information is encoded in the frequency domain and the signal may be of bounded amplitude, the required power will be independent of the number of qubits represented. By contrast, although sophisticated digital simulations of up to 40 qubits have been performed on large, high-performance computers, they are bulky and require several megawatts of power to operate [28].

The ability to emulate a quantum computer classically comes with certain additional freedoms and restrictions. Aaronson has shown, for example, that a quantum computer with access to a history of hidden-variables is capable of performing an unstructured search faster than Grover's algorithm [1]. However, a classical emulation is not necessarily restricted to unitary operations on the state. In particular, projections and inner products may be performed explicitly, allowing one to extract components of the quantum state that would otherwise be inaccessible via unitary evolution and measurement alone. This confers a significant computational advantage in terms of the number of operations needed to perform a search. Since, in our QMT representation, the Hilbert space is embedded in the frequency domain of the signal, bandwidth becomes the primary limitation. Nevertheless, quite large problems (on the order of 40 qubits) could be represented by this scheme, and hybrid schemes may be used to solve larger problems while still taking advantage of the inherent features of quantum parallelism and superposition [8].

In an effort to demonstrate the feasibility of this approach, we have developed a fully programmable hardware prototype device capable of emulating a two-qubit quantum computer implementing a universal set of gate operations with over 99% fidelity [16]. Scaling to larger numbers of qubits would require bandwidths and physical components that scale exponentially by roughly a factor of 1000 for every 10 additional qubits. Exisiting semiconductor manufacturing technologies can place over a billion transistors on a single integrated circuit chip, which is sufficient to emulate about 30 qubits using our approach. Future manufacturing capabilities are expected to extend this to perhaps 40 qubits within the next decade. Unlike traditional analog computers, a quantum emulation device would also be capable, if need be, of implementing standard quantum error correction protocols, albeit at the cost of additional qubits and, hence, bandwidth, so that fault-tolerant operation may, in principle, be possible.

This paper will examine the relative advantages and trade-offs for such an approach as compared to either classical brute-force searches or applications of Grover's algorithm. Since gate fidelity will be important in determining the efficacy of each approach, we consider the inclusion of additive white complex Gaussian noise to the emulating signal, which reproduces the effect of a depolarizing channel. Performance is then assessed as a function of the overall fidelity or, equivalently, the signal-to-noise ratio (SNR).

The organization of the paper is as follows. In Sect. 2, we introduce the general subspace projection method used to solve the Boolean function problem for one solution and multiple solutions. Section 3 defines a general noisy quantum oracle that may be used to represent any given Boolean function in terms of its solutions. An assessment of relative performance is described and analyzed in Sect. 4, and our conclusions are summarized in Sect. 5.

## 2 Subspace projection method

### 2.1 QMT representation

In the QMT representation of Ref. [17], each qubit is represented by a complex exponential signal of a given frequency. For $n + 1$ qubits, let $\omega_0, \ldots, \omega_n$ denote the qubit angular frequencies, which are taken to be octavely spaced so that $\omega_k = 2^k \omega_0$ for some minimum frequency $\omega_0 > 0$. Thus, qubits $|0\rangle_k$ and $|1\rangle_k$ are represented by the time-dependent signals $e^{+i\omega_k t}$ and $e^{-i\omega_k t}$, respectively. A given computational basis function $|x, y\rangle$, where $x \in \{0, \ldots, N-1\}$ and $y \in \{0, 1\}$ represent the input and output registers, respectively, is given by the product of the $n + 1$ constituent qubit signals in accordance with the binary expansion of $x$. Thus, a given quantum state $|\psi\rangle$ with components $\alpha_{x,y} \in \mathbb{C}$ is represented by

$$\psi(t) = \sum_{x=0}^{N-1} \sum_{y=0}^{1} \alpha_{x,y} e^{i\Omega_{x,y}t}, \tag{1}$$

where $\Omega_{x,y} = (2N - 1 - 4x - 2y)\omega_0$. The state $|\psi\rangle$ is taken to be normalized to a magnitude of $s > 0$ so that

$$\sum_{x=0}^{N-1} \sum_{y=0}^{1} |\alpha_{x,y}|^2 = s^2. \tag{2}$$

Finally, the inner product between any two such signals $\phi$ and $\psi$ is defined to be

$$\langle \phi | \psi \rangle = \frac{1}{T} \int_0^T \phi(t)^* \psi(t) \, dt, \tag{3}$$

where $T \in 2\pi \mathbb{N}/\Omega_0$ is an integer number of periods. Note, in particular, that this implies $\alpha_{x,y} = \langle x, y | \psi \rangle$. Physically, the inner product may be obtained by low-pass filtering the real and imaginary components of the complex product $\phi^*(t)\psi(t)$, which may be obtained through the use of four-quadrant multipliers and operational amplifiers.

In addition to the usual unitary gate operations, the QMT representation also allows for the construction of projections onto signal subspaces through the use of specialized bandpass filters. This provides a unique capability to "disentangle" entangled (i.e., nonseparable) states and is the primary mechanism upon which the subspace method of solution is based. In the following sections, we describe how this method may be used to solve Boolean problems of one or more solutions.

### 2.2 Problems with one solution

Suppose we have an oracle function $f : \{0, \ldots, N-1\} \mapsto \{0, 1\}$, where $N = 2^n$, such that $f(x) = 1$ if $x = a$ and $f(x) = 0$ otherwise. Let $U_f$ denote a unitary transformation corresponding to $f$ and such that, for a given basis state $|x, y\rangle$, where $x \in \{0, \ldots, N-1\}$ and $y \in \{0, 1\}$, we have

$$U_f |x, y\rangle = |x, y \oplus f(x)\rangle, \tag{4}$$

where $\oplus$ is the modulo-2 binary sum operation.

We begin with an initial state of the form $s|0, 0\rangle$ and apply Hadamard gates $\mathbf{H}_1, \ldots, \mathbf{H}_n$ to each qubit in the input register to obtain

$$|\psi_0\rangle = \mathbf{H}_n \cdots \mathbf{H}_1 s|0, 0\rangle = \frac{s}{\sqrt{N}} \sum_{x=0}^{N-1} |x, 0\rangle. \tag{5}$$

Next, we apply $U_f$ to $|\psi_0\rangle$ to obtain

$$|\psi'\rangle = \frac{s}{\sqrt{N}} \sum_{x=0}^{N-1} |x, f(x)\rangle = \frac{s}{\sqrt{N}} \left[ \sum_{x \neq a} |x, 0\rangle + |a, 1\rangle \right]. \tag{6}$$

In a true quantum system, a measurement of the output register would result in the solution $|a, 1\rangle$ only with a vanishingly small probability of $1/N$. Using our classical signal-based representation, one may perform a projection onto the subspace of qubit 0 corresponding to the output register value of $|1\rangle$, denoted $|\psi\rangle = \Pi_1^{(0)}|\psi'\rangle$, to obtain

$$|\psi\rangle = \Pi_1^{(0)}|\psi'\rangle = \frac{s}{\sqrt{N}}|a, 1\rangle. \tag{7}$$

The solution $a$ is thereupon read off from the value of the input register using $n$ single-qubit measurement gates.

Of course, even in a classical QMT representation, the solution component of the signal has an amplitude that is $N - 1$ times smaller than that of the nonsolution component, so discrimination between the two remains a similar challenge. The question we ask is whether, in a practical setting, a classical subspace projection approach may provide computational gain over other classical or quantum alternatives. Before examining this question, however, we turn to a generalization of the single-solution problem.

### 2.3 Problems with multiple solutions

Consider an $n$-bit Boolean function $f$ such that, for $x \in \{0, \ldots, N-1\}$ and $M \in \{0, \ldots, N\}$, we have $f(x) = 1$ if $x \in \{a_1, \ldots, a_M\} = \mathcal{S}$ and $f(x) = 0$ otherwise. Each element of $\mathcal{S}$ is assumed unique, so $a_j = a_{j'}$ only if $j = j'$. We also consider the possibility that $M = 0$, in which case $f(x) = 0$ for all $x$. As before, let $U_f$ denote the corresponding unitary transformation.

We begin, as before, with the initial $(n+1)$-qubit state $s|0, 0\rangle$ and apply Hadamard gates to the input register to obtain $|\psi_0\rangle$, then apply the oracle $U_f$ to obtain

$$|\psi'\rangle = \frac{s}{\sqrt{N}} \left[ \sum_{x \notin \mathcal{S}} |x, 0\rangle + \sum_{j=1}^{M} |a_j, 1\rangle \right]. \tag{8}$$

The solution subspace is then given by the projection

$$|\psi\rangle = \Pi_1^{(0)}|\psi'\rangle = \frac{s}{\sqrt{N}} \sum_{j=1}^{M} |a_j, 1\rangle. \tag{9}$$

A measurement of this state will provide one of the $M$ solutions. If the number of solutions is known, the procedure of preparation, projection, and measurement may be repeated $\mathcal{O}(M)$ times to obtain all $M$ solutions.

### 2.4 Finding the number of solutions

One may also determine the number of solutions from the projected state. To do this, consider an auxiliary state of the form

$$|\phi\rangle = \frac{s}{\sqrt{N}} \sum_{x=0}^{N-1} |x, 1\rangle. \tag{10}$$

A representative signal of this form is easily constructed as a product of $n$ cosines, with frequencies $\omega_1, \ldots, \omega_n$ and a complex exponential of the form $e^{-i\omega_0 t}$. The inner product of the state $|\phi\rangle$ with the projection state $|\psi\rangle$ gives the number of solutions, since

$$M = \frac{N}{s^2} \langle \phi | \psi \rangle. \tag{11}$$

One may now perform a measurement on the input register of $|\psi\rangle$, resulting in an outcome $x = a_j$, from which one may construct the solution vector

$$|\phi_j\rangle = \frac{s}{\sqrt{N}} |a_j, 1\rangle \tag{12}$$

The state $|\phi_j\rangle$ can now be subtracted from $|\psi\rangle$, and one may continue in this manner until all $M$ solutions have been extracted.

## 3 General quantum oracles

In order to realize the algorithm for a particular instance of the Boolean function $f$, we must construct a corresponding quantum oracle. The details of how such an oracle may be constructed from more elementary gate operations are problem specific and typically based on a reversible-logic formulation of the Boolean function itself. Here we will abstract the details of any particular oracle and consider instead a template construction for an arbitrary oracle given a set of solutions as defining parameters. We note that the use of quantum oracles with planted solutions is done solely for purposes of analysis; no knowledge of the solutions is used by the algorithms themselves.

### 3.1 Ideal quantum oracles

First of all, if there are no solutions (i.e., $\mathcal{S} = \emptyset$), then $U_f$ is trivially equivalent to the identity. Suppose instead that there is exactly one solution, given by an $n$-bit integer $a \in \{0, \ldots, N-1\}$ with a little-endian binary representation such that $a = [a]_{n-1}2^{n-1} + \cdots + [a]_0 2^0$. The corresponding Boolean function is implemented by the unitary operator

$$U_f = \mathbf{A}\,\mathbf{C}_{n\cdots 1,0}\,\mathbf{A}, \tag{13}$$

where $\mathbf{C}_{n\cdots 1,0}$ is an $n$-fold Toffoli gate with control qubits 1 through $n$ and target qubit 0. The operator $\mathbf{A}$ is defined by the parameter $a$ such that

$$\mathbf{A} = \prod_{i=0}^{n-1} \mathbf{X}_{i+1}^{1-[a]_i}, \tag{14}$$

where the exponent $1 - [a]_i$ determines whether to apply the corresponding NOT gate that swaps qubits $|0\rangle$ and $|1\rangle$.

A standard construction allows the $n$-fold Toffoli gate to be decomposed into $48n^2 + \mathcal{O}(n)$ one- and two-qubit gate operations, although the inclusion of at least one ancilla qubit can reduce this size to $\mathcal{O}(n)$ [4]. A QMT representation of the $n$-fold Toffoli gate may be realized more simply by applying narrowband filters to the frequencies $\Omega_{N-1,0}$ and $\Omega_{N-1,1}$ and then physically swapping the coefficients $\alpha_{N-1,0}$ and $\alpha_{N-1,1}$, represented by the DC filter outputs, to obtain

$$\mathbf{C}_{n\cdots 1,0}|\psi\rangle = \alpha_{N-1,1}|N-1,0\rangle + \alpha_{N-1,0}|N-1,1\rangle + \sum_{x=0}^{N-2}\sum_{y=0}^{1} \alpha_{x,y}|x,y\rangle. \tag{15}$$

For multiple solutions $a_1, \ldots, a_M \in \{0, \ldots, N-1\}$ we may define similarly the corresponding operators $\mathbf{A}_1, \ldots, \mathbf{A}_M$. Provided the solutions are unique, the oracle is then given by

$$U_f = \prod_{j=1}^{M} \left[\mathbf{A}_j\,\mathbf{C}_{n\cdots 1,0}\,\mathbf{A}_j\right]. \tag{16}$$

where

$$\mathbf{A}_j = \prod_{i=0}^{n-1} \mathbf{X}_{i+1}^{1-[a_j]_i}. \tag{17}$$

### 3.2 Noisy oracles

There are many ways one might model imperfect gates. For simplicity, let us suppose that errors occur only in the oracle and that each application of the oracle results in a transformation $\tilde{U}_f$ such that, for a given state $|\psi\rangle$,

$$\tilde{U}_f|\psi\rangle = U_f|\psi\rangle + |\nu\rangle, \tag{18}$$

where $|\nu\rangle$ is a random state to be defined below. Note that, since $\tilde{U}_f|\psi\rangle$ defines an ensemble of pure states, we may think of it as representing a mixed state.

We now turn to defining $|\nu\rangle$. Let $w$ denote an additive white noise complex Gaussian process with power spectral density $\sigma^2$ such that $E[w(t)] = 0$ and $E[w(t')^* w(t)] = \sigma^2 \delta(t' - t)$. Filtering onto the $2N$ component frequencies forms a projected state $|\nu\rangle$ of the form

$$|\nu\rangle = \sum_{x=0}^{N-1} \sum_{y=0}^{1} |x, y\rangle \langle x, y|w\rangle = \sqrt{\frac{\sigma^2}{T}} \sum_{x=0}^{N-1} \sum_{y=0}^{1} z_{x,y}|x, y\rangle, \quad (19)$$

where $z_{x,y}$ are complex Gaussian random variables with $E[z_{x,y}] = 0$ and $E[z_{x',y'}^* z_{x,y}] = \delta_{x,x'} \delta_{y,y'}$. Thus, the error incurred by application of the noisy oracle corresponds to a depolarizing channel.

The fidelity of the oracle may be determined as follows. Recall that $\tilde{U}_f|\psi\rangle$ defines a mixed state, which we shall denote $\rho$. Since the components of $|\nu\rangle$ form a set of uncorrelated complex Gaussian random variables, the unnormalized form of $\rho$ is given by

$$\rho = |U_f \psi\rangle \langle U_f \psi| + \frac{\sigma^2}{T} \mathbf{I}, \quad (20)$$

where $\mathbf{I}$ is the $2N$-dimensional identity operator [17]. The fidelity $F$ is therefore given by

$$F^2 = \frac{\langle U_f \psi | \rho | U_f \psi \rangle}{\|U_f \psi\|^2 \operatorname{Tr}[\rho]} = \frac{s^2 + \sigma^2/T}{s^2 + 2N\sigma^2/T}. \quad (21)$$

Note that $F$ depends only on $N$ and the SNR value $S^2 = s^2 T/\sigma^2$. Equivalently, we may specify the value of $\sigma^2$ needed for a given $F$ to be

$$\sigma^2 = \frac{s^2 T (1 - F^2)}{2N F^2 - 1}. \quad (22)$$

Note that there is no value of $\sigma^2$ high enough to give an oracle fidelity of $F \leq 1/\sqrt{2N}$. Although characterized in terms of additive noise, the oracle of Eq. (20) may be considered a description of a general depolarizing channel for a true quantum system exhibiting the same level of fidelity.

### 3.3 Noisy solution estimates

As an illustration, consider estimating the number of solutions using the subspace projection method. For a noisy oracle, the number of solutions may be estimated via the subspace projection method using the random complex quantity

$$\tilde{M} = \frac{N}{s^2}\left\langle\phi\left|\Pi_1^{(0)}\tilde{U}_f\right|\psi_0\right\rangle = M + \frac{N}{s^2}\langle\phi|\nu\rangle. \quad (23)$$

Note that $\tilde{M}$ has a complex Gaussian distribution with mean $M$ and variance $N^2\sigma^2/(s^2T)$. Thus, $|\tilde{M}|$ has a Rician distribution with mean

$$E\left[|\tilde{M}|\right] = \frac{N\sigma}{2s}\sqrt{\frac{\pi}{T}}L_{1/2}\left(-\frac{M^2s^2T}{N^2\sigma^2}\right) \quad (24)$$

and variance

$$\text{Var}\left[|\tilde{M}|\right] = \frac{N^2\sigma^2}{s^2T} + M^2 - E\left[|\tilde{M}|\right]^2, \quad (25)$$

where $L_{1/2}$ is the Laguerre polynomial of degree $1/2$.

The number of solutions, which must of course be an integer, may be estimated by

$$\hat{M} = \left\lfloor |\tilde{M}| + \frac{1}{2} \right\rfloor. \quad (26)$$

The probability of obtaining a given value $m$ from this estimate is therefore

$$\Pr\left[\hat{M} = m\right] = \Pr\left[m - \frac{1}{2} \leq |\tilde{M}| < m + \frac{1}{2}\right], \quad (27)$$

which may be written in terms of the Marcum Q-function using the cumulative distribution function of the Rician random variable $|\tilde{M}|$ [21].

## 4 Comparison of methods

Ideally, the subspace projection method always gives a correct answer with a single application of the oracle. This is in contrast to a brute-force search, which requires $\mathcal{O}(N)$ oracle calls, or Grover's search algorithm, which gives the correct answer with an error of $\mathcal{O}(1/N)$ using $\mathcal{O}(\sqrt{N})$ oracle calls. Imperfect gate operators will, however, degrade search performance. In this section, we will use the noisy oracle model to compare the performance of the subspace projection method to that of both Grover's algorithm and a simple, brute-force search when the number of solutions is known.

### 4.1 Brute-Force approach

Suppose we prepare the state $|\psi_0\rangle$ given by Eq. (5), apply the noisy oracle to obtain $|\tilde{\psi}\rangle = \tilde{U}_f|\psi_0\rangle$, and perform a measurement on all $n+1$ qubits of $|\tilde{\psi}\rangle$ to obtain an outcome $(x, y)$. What is the probability that $(x, y) = (a, 1)$? If we use $|\tilde{\psi}\rangle$ to determine probabilities according to the Born rule and use, say, an independent, uniformly

distributed random variable $u \in [0, 1]$ to determine the outcome, then the probability of success for a particular realization is

$$\tilde{p}_B = \Pr\left[u \leq \frac{|\langle a, 1|\tilde{\psi}\rangle|^2}{\|\tilde{\psi}\|^2}\right] = \frac{|\langle a, 1|\tilde{\psi}\rangle|^2}{\|\tilde{\psi}\|^2}. \tag{28}$$

As described previously, the ensemble of all realizations of $|\tilde{\psi}\rangle$ is represented by a mixed state $\rho$, where

$$\rho = \frac{s^2}{N} \sum_{x=0}^{N-1} \sum_{x'=0}^{N-1} |x, f(x)\rangle\langle x', f(x')| + \frac{\sigma^2}{T}\mathsf{I}. \tag{29}$$

Over this ensemble, then, the probability of success is

$$p_B = E[\tilde{p}_B] = \frac{\langle a, 1|\rho|a, 1\rangle}{\mathrm{Tr}[\rho]} = \frac{s^2/N + \sigma^2/T}{s^2 + 2N\sigma^2/T}, \tag{30}$$

since

$$\langle a, 1|\rho|a, 1\rangle = \frac{s^2}{N} + \frac{\sigma^2}{T} \tag{31}$$

and

$$\mathrm{Tr}[\rho] = s^2 + \frac{2N\sigma^2}{T}. \tag{32}$$

Note that $p_B = 1/N$ for $\sigma = 0$, as expected, and $p_B$ goes as $1/(2N)$ for $s^2 T/(N\sigma^2) \ll 1$.

The success probability for any one trial is, of course, quite low. However, the system can be reprepared and measured $N$ times to yield a total probability for at least one success of

$$P_B = 1 - (1 - p_B)^N. \tag{33}$$

For large $N$, and keeping all other parameters fixed, $P_B$ tends to the value $1 - e^{-1/2} \approx 0.3935$.

For multiple solutions $a_1, \ldots, a_M$, the probability of success for a single trial is simply

$$p_B = \sum_{j=1}^{M} \frac{\langle a_j, 1|\rho|a_j, 1\rangle}{\mathrm{Tr}[\rho]} = \frac{s^2 M/N + M\sigma^2/T}{s^2 + 2N\sigma^2/T}. \tag{34}$$

Note that, in the extreme case of $M = N$, this probability tends to $1/2$ for $N\sigma^2/(s^2 T) \gg 1$, indicating the effect of noise randomly flipping the output register qubit. If there are no solutions (i.e., $M = 0$), then $p_B = 0$, as expected.

### 4.2 Subspace projection method

The subspace projection method is similar to the brute-force approach except that a final projection operator $\Pi_1^{(0)}$ is applied prior to measurement, resulting in a random state of the form $|\tilde{\psi}\rangle = \Pi_1^{(0)} \tilde{U}_f |\psi_0\rangle$. (We assume, for simplicity, that the projection operator itself is not noisy.) For a single solution, the expected probability of success over an ensemble of noisy oracles is

$$p_S = \frac{\langle a, 1 | \Pi_1^{(0)} \rho \Pi_1^{(0)} | a, 1 \rangle}{\text{Tr}\left[\Pi_1^{(0)} \rho \Pi_1^{(0)}\right]}, \tag{35}$$

where $\rho$ is given by Eq. (29).

Since

$$\langle a, 1 | \Pi_1^{(0)} \rho \Pi_1^{(0)} | a, 1 \rangle = \frac{s^2}{N} + \frac{\sigma^2}{T} \tag{36}$$

and

$$\text{Tr}\left[\Pi_1^{(0)} \rho \Pi_1^{(0)}\right] = \frac{s^2}{N} + \frac{N\sigma^2}{T}, \tag{37}$$

we conclude that

$$p_S = \frac{s^2/N + \sigma^2/T}{s^2/N + N\sigma^2/T}. \tag{38}$$

Note that, for high SNR, $p_S$ tends to unity, as expected, while for low SNR it goes as $1/N$. The latter result indicates that the probability of success falls off twice as slowly as that of the brute-force approach as a consequence of the fact that the output register has been projected onto the $|1\rangle$ state.

For multiple solutions, the probability of success for $M \geq 1$ is

$$p_S = \sum_{j=1}^{M} \frac{\langle a_j, 1 | \Pi_1^{(0)} \rho \Pi_1^{(0)} | a_j, 1 \rangle}{\text{Tr}\left[\Pi_1^{(0)} \rho \Pi_1^{(0)}\right]} = \frac{s^2 M/N + M\sigma^2/T}{s^2 M/N + N\sigma^2/T}. \tag{39}$$

When $M = N$, Eq. (39) correctly predicts the method will always produce a correct solution. By contrast, the brute-force approach produces a correct solution only half of the time. To understand this, note that, in the case of a very noisy oracle, roughly half of the $N$ solutions will have the output qubit flipped. Nevertheless, the subspace with $y = 1$ onto which we project constitutes about half of the valid solutions.

The case of no solutions (i.e., $M = 0$) is special. If $\sigma > 0$, then the subspace for which $y = 1$ will not be entirely empty, due to noise, and so there is a nonzero probability of obtaining a solution, albeit an erroneous one. Thus, the probability of obtaining a correct solution is zero, as predicted by Eq. (39). If there is no noise (i.e., $\sigma = 0$), then $\Pi_1^{(0)} \rho \Pi_1^{(0)}$ is the zero operator and no longer constitutes a normalizable mixed state. For this special case ($M = 0, \sigma = 0$), we simply *define* $p_S$ to be zero, as this is the limiting value of $p_S$ as $\sigma \to 0$.

In comparing the subspace projection method to the brute-force approach, we readily observe that, for all parameter values,

$$p_B \leq p_S. \tag{40}$$

This result is not surprising, as the subspace projection method provides additional refinement of the state by projecting onto the solution space. In particular, if the subspace projection method is applied $N$ times, we will have a total probability of at least one success of

$$P_S = 1 - (1 - p_S)^N, \tag{41}$$

which tends to a limiting value of $1 - e^{-1} \approx 0.6321$ for large $N$, provided $p_S > 0$ and all other parameters are held fixed.

### 4.3 Grover's search algorithm

In Grover's search algorithm, one begins with a state of the form $|\psi_0\rangle$, as given by Eq. (5), transforms the output register to obtain $\mathbf{H}_0 \mathbf{X}_0$, and then applies $R$ iterations of the (ideal) Grover operator $\mathbf{G}$ and a final $\mathbf{H}_0$ to obtain, for a single solution, the final state

$$\begin{aligned}|\psi_R\rangle &= \mathbf{H}_0 \mathbf{G}^R \mathbf{H}_0 \mathbf{X}_0 |\psi_0\rangle \\ &= s \sin(R\theta + \theta/2) |a, 1\rangle + \frac{s \cos(R\theta + \theta/2)}{\sqrt{N-1}} \sum_{x \neq a} |x, 1\rangle,\end{aligned} \tag{42}$$

where $\sin(\theta/2) = 1/\sqrt{N}$. For optimality, the value $R = \lfloor \pi \sqrt{N}/4 \rfloor$ is chosen.

The Grover operator is composed of the oracle operator $U_f$ and a diffusion operator $\mathbf{W}$ so that $\mathbf{G} = \mathbf{W} U_f$. In the case of a noisy oracle, the Grover operator becomes $\tilde{\mathbf{G}} = \mathbf{W} \tilde{U}_f$. (We assume, for simplicity, that the diffusion operator is not noisy.) Operating on a given state $|\psi\rangle$, then, gives $\mathbf{G}|\psi\rangle + \mathbf{W}|\nu\rangle$. Since $\mathbf{W}$ is unitary and $|\nu\rangle$ has multivariate Gaussian components, the distribution of $\mathbf{W}|\nu\rangle$ will be the same as that of $|\nu\rangle$. Thus, we shall write $\tilde{\mathbf{G}}|\psi\rangle = \mathbf{G}|\psi\rangle + |\nu_1\rangle$ and, consequently, $\tilde{\mathbf{G}}^R |\psi\rangle = \mathbf{G}^R |\psi\rangle + |\nu_1\rangle + \cdots + |\nu_R\rangle$, where $|\nu_1\rangle, \ldots, |\nu_R\rangle$ are independent and identically distributed as $|\nu\rangle$ of Eq. (19). Thus, using a noisy oracle we obtain a final state

$$|\tilde{\psi}_R\rangle = \mathbf{H}_0 \tilde{\mathbf{G}}^R \mathbf{H}_0 \mathbf{X}_0 |\psi_0\rangle$$
$$= s \sin(R\theta + \theta/2)|a, 1\rangle + \frac{s \cos(R\theta + \theta/2)}{\sqrt{N-1}} \sum_{x \neq a} |x, 0\rangle + \sum_{r=1}^{R} |v_r\rangle, \quad (43)$$

As before, the ensemble of pure states $|\tilde{\psi}_R\rangle$ constitutes a mixed state, this one of the form

$$\rho_R = |\psi_R\rangle\langle\psi_R| + \frac{R\sigma^2}{T}\mathbf{I}. \quad (44)$$

The expected probability of success is therefore

$$p_G = \frac{\langle a, 1|\rho_R|a, 1\rangle}{\mathrm{Tr}[\rho_R]} = \frac{s^2 \sin^2(R\theta + \theta/2) + R\sigma^2/T}{s^2 + 2NR\sigma^2/T}, \quad (45)$$

since

$$\langle a, 1|\rho_R|a, 1\rangle = s^2 \sin^2(R\theta + \theta/2) + \frac{R\sigma^2}{T} \quad (46)$$

and

$$\mathrm{Tr}[\rho_R] = \sum_{x=0}^{N-1} \sum_{y=0}^{1} |\langle x, y|\psi_R\rangle|^2 + \mathrm{Tr}\left[\frac{R\sigma^2}{T}\mathbf{I}\right]$$
$$= s^2 \sin^2(R\theta + \theta/2) + s^2 \cos^2(R\theta + \theta/2) + \frac{2NR\sigma^2}{T} \quad (47)$$
$$= s^2 + \frac{2NR\sigma^2}{T}.$$

For multiple solutions, the Grover algorithm must be slightly modified. For $1 \leq M \leq N/2$ solutions, the optimal number of iterations is $R = \lfloor \pi \sqrt{N/M}/4 \rfloor$. The case of $M > N/2$ can be handled by simply redefining the oracle such that $f(x)$ is replaced by $1 - f(x)$, in which case $R = \lfloor \pi \sqrt{(N-M)/M}/4 \rfloor$ is optimal. Otherwise, the procedure is the same, and the final state after $R$ iterations of a noisy Grover operator is

$$|\tilde{\psi}_R\rangle = \frac{s \sin(R\theta + \theta/2)}{\sqrt{M}} \sum_{j=1}^{M} |a_j, 1\rangle + \frac{s \cos(R\theta + \theta/2)}{\sqrt{N-M}} \sum_{x \notin S} |x, 0\rangle + \sum_{r=1}^{R} |v_r\rangle, (48)$$

where $\sin(\theta/2) = \sqrt{M/N}$. The expected probability of finding at least one solution is therefore

$$p_G = \frac{s^2 \sin^2(R\theta + \theta/2) + MR\sigma^2/T}{s^2 + 2NR\sigma^2/T}. \quad (49)$$

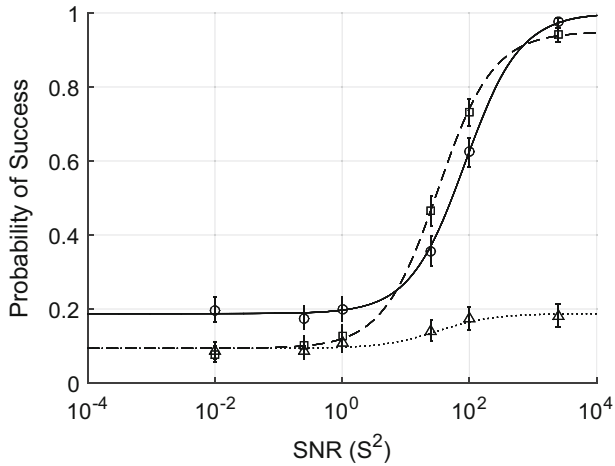

**Fig. 1** Plot of the probability of success versus SNR ($S^2$) for $N = 16$ and $M = 3$ over the three solution methods: the subspace projection method (*solid line*, *circles*), Grover's algorithm (*dashed line*, *squares*), and the brute-force approach (*dotted line*, *triangles*). The *error bars* indicate the (exact) Clopper–Pearson 95% confidence intervals for 1000 realizations

In the special case of no solutions (i.e., $M = 0$), then $\theta = 0$ and the number of iterations is taken to be zero as well (i.e., $R = 0$). Thus, Eq. (49) correctly predicts a probability of zero for obtaining a correct solution.

Figure 1 shows the expected probability of success for the three methods as a function of SNR for parameter values of $N = 16$ and $M = 3$. Using a simulation of the QMT representation with ideal filters and a noisy oracle, the three methods were applied to 1000 instances of the same problem for several different SNR values. In all cases, good agreement was found between the numerical results and the theoretical expectation value, thereby validating both.

Unlike that of $p_B$ and $p_S$, we note that there is no simple relationship between $p_G$ and $p_S$. For some parameter values $p_G > p_S$, for others $p_G < p_S$, and in certain special cases (e.g., $M = 0$ or $M = N$) the two are identical. As the SNR tends to zero ($S^2 \to 0$), $p_S$ tends to $M/N$, whereas $p_G$ tends to half this value. As the SNR becomes large ($S^2 \gg 1$), we find that $p_S$ tends to unity, as expected, whereas $p_G$ tends to a possibly smaller limiting value of $\sin^2(R\theta + \theta/2)$. Thus, $p_S \geq p_G$ in these two limits. For intermediate SNR values, it is possible that $p_S < p_G$.

Comparing $p_S$ and $p_G$ and writing $S^2 = s^2 T/\sigma^2$ for the SNR, we see that $p_S > p_G$ if and only if

$$M \cos^2(R\theta + \theta/2) S^4 - bS^2 + MN^2 R > 0, \qquad (50)$$

where

$$b = N^2 \sin^2(R\theta + \theta/2) - M(2NR + N - MR). \qquad (51)$$

When $M \cos^2(R\theta + \theta/2) = 0$ we have a linear inequality in $S^2$ such that $p_S > p_G$ for $S^2 < S_0^2 = MN^2 R/b$ and $p_S < p_G$ for $S^2 > S_0^2$. In particular, $\cos(R\theta + \theta/2) = 0$

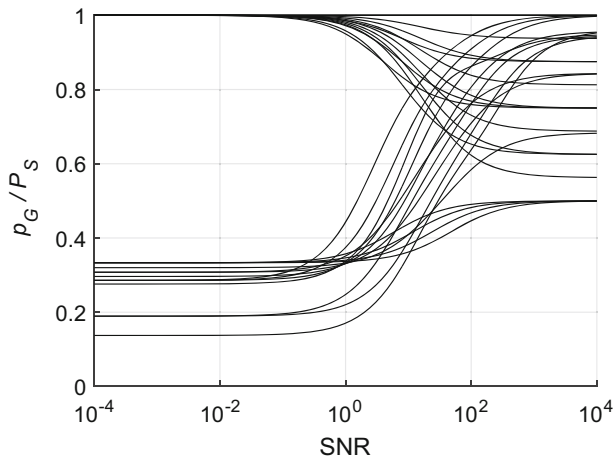

**Fig. 2** Plot of the ratio $p_G/P_S$, the probability of success using Grover's algorithm and the subspace method using the same number of oracle calls, versus the signal-to-noise ratio $S^2$ for $n \leq 4$ and $M \leq 2^n$. Note that all *curves* around bounded from above by one. Larger values of $n$ and $M$ produce *curves* with similar behavior but are omitted for simplicity

whenever $M = N/4$, since, in this case, $R = 1$ and $\theta = \pi/3$. Note that $S_0$ is undefined for $M = 0$ and $M = N$, since $b = 0$; however, in each of these cases $p_S = p_G$ for all $S$.

For $1 \leq M < N/4$, we find that $p_S < p_G$ only for the intermediate values $S_-^2 < S^2 < S_+^2$, where

$$S_\pm^2 = \frac{b \pm \sqrt{b^2 - 4M^2 N^2 R \cos^2(R\theta + \theta/2)}}{2M \cos^2(R\theta + \theta/2)}. \tag{52}$$

It may also be observed that $S_- > 1$, so $p_S < p_G$ only for over-unity SNR values. Finally, for $N/4 < M < N$ we have $p_S > p_G$ for all $S^2$.

Of course, the subspace method and Grover's algorithm differ significantly in the required number of oracle calls. The former requires one, whereas the latter requires $R$. Suppose, though, the subspace method is repeated $R + 1$ times. In this case, the overall probability of success (i.e., of obtaining at least one valid solution) is

$$P_S = 1 - (1 - p_S)^{R+1}. \tag{53}$$

Figure 2 shows a plot of the ratio $p_G/P_S$ for $n \leq 4$ and $M \leq 2^n$. Based on this and similar numerical analyses (not shown here), it may be conjectured that $P_S \geq p_G$ for all $n$, $M$, and $S$. Thus, for the same computational effort, the subspace method appears to provide equal or better performance to Grover's algorithm. This result does not, however, conflict with the well-established optimality of Grover's algorithm, as the latter is based on a constraint of unitary gate operations, one to which the proposed classical representation is not bound [5].

While a comparison of the subspace projection method to Grover's algorithm is relatively straightforward, a similar comparison to a digital solver proves somewhat more challenging. We may imagine that a digital device performs an ideal brute-force search with near-infinite SNR, giving a probability of success of $p_B = M/N$ for any particular instance. To place this approach on the same footing as that of the other two methods, we may restrict the number of oracle calls to $R + 1$ so that the probability of obtaining at least one valid solution is $1 - (1 - M/N)^{R+1}$. Since $p_S \geq M/N$, this is still no better than the subspace projection method and will be better than Grover's algorithm only for low-SNR oracles.

More challenging still is a comparison in terms of absolute solution time. Current digital processors operate at near GHz clock speeds, giving each fundamental logic operation an execution time on the order of a nanosecond. Memory access times can extend this time to tens of nanoseconds. A quantum emulation device implementing the subspace projection algorithm with, say, 10 qubits in the frequency range of 1 MHz to 1 GHz would have a gate operation time on the order of a microsecond but would evaluate a thousand digital inputs at once, giving it an effective speed at least equivalent to that of the digital processor. Thus, it is reasonable to suppose that the algorithmic advantage observed in terms of the number of required oracle calls would indeed translate into an actual speed advantage for computation.

## 5 Conclusion

This paper has introduced a novel approach to performing unstructured searches using a classical emulation of a gate-based quantum computer. In this approach, the classical, signal-based representation of quantum states allows for a direct computation of subspace projection operations, thereby allowing additional computational capability beyond the application of unitary gate operations. The price paid for this capability is a limitation in scale, due to bandwidth constraints, but this by no means nullifies the utility of the approach.

Under ideal conditions, the proposed subspace projection method is capable of finding a solution in an unstructured search using a single oracle call, independent of the size of the problem. Under more realistic conditions, however, the approach will be hampered by noise and other imperfections such that a solution may be found only with a certain probability of success. The computation may then be repeated to improve the overall success probability. The number of solutions may also be estimated by this method, and, unlike Grover's algorithm, the probability of success will not be sensitive to this estimate.

To understand the efficacy of this approach in a realistic setting, we have compared the probability of success of the proposed subspace projection method to that of a simple, brute-force search or application of Grover's search algorithm, modeling the imperfections as a noisy oracle. In order to facilitate comparison across the three methods, we have ignored errors in both the projection and Grover diffusion operations under the simplifying assumption that the oracle is the dominant source of error. For an unstructured search, a single instance of the subspace projection method always provides a better probability of success, not surprisingly, than the brute-force approach using a single oracle call. A single instance of Grover's algorithm tends to provide a higher probability of success for intermediate values of the noisy oracle's signal-to-noise ratio, albeit with many more oracle calls. Interestingly, when repeated instances of the subspace projection method are considered, it was found that for the same total number of oracle calls the subspace projection method provides a higher probability of success than does Grover's algorithm.

**Acknowledgements** This work was support by the Office of Naval Research under Grant No. N00014-14-1-0323.